\documentclass[conference,10pt]{IEEEtran}
\IEEEoverridecommandlockouts

\usepackage{cite}
\ifCLASSINFOpdf
  \usepackage[pdftex]{graphicx}
\else
   \usepackage[dvips]{graphicx}
\fi
\usepackage{amsmath,bm}
\usepackage{amssymb}
\DeclareMathOperator*{\minimize}{minimize}

\ifCLASSOPTIONcompsoc
  \usepackage[caption=false,font=normalsize,labelfont=sf,textfont=sf]{subfig}
\else
  \usepackage[caption=false,font=footnotesize]{subfig}
\fi

\usepackage{stfloats}
\usepackage{url}
\usepackage{multirow}
\usepackage{graphicx}

\usepackage[printonlyused,nohyperlinks,nolist]{acronym}
\usepackage{algorithm}
\usepackage{algpseudocode}

\usepackage{dsfont}

\usepackage{siunitx}
\usepackage{import}
\usepackage{color}

\DeclareUnicodeCharacter{2212}{-}

\def\BibTeX{{\rm B\kern-.05em{\sc i\kern-.025em b}\kern-.08em
    T\kern-.1667em\lower.7ex\hbox{E}\kern-.125emX}}

\usepackage{amsthm}
\newtheorem{remark}{Remark}
\newtheorem{proposition}{Proposition}

\usepackage{scalefnt}

\usepackage[top=0.75in, bottom=1in, left=0.625in, right=0.625in]{geometry}

\begin{document}

%------------------------------------------------------------------
% Acronyms
% use:  \ac{BS}         to use a acronym (first time = full name, then only acronym)
%       \acp{BS}        use the plural
%       \acf{BS}        print the full name and ignore previous declaration
%       \acs{BS}        Use the acronym, even before the first corresponding \ac command
%       \acl{acronym}   Expand the acronym without using the acronym itself.
%       \acresetall     Reset all acronyms (useful after abstract)
%------------------------------------------------------------------
\begin{acronym}
    \acro{los}[LoS]{line-of-sight}
    \acro{ul}[UL]{uplink}
    \acro{dl}[DL]{downlink}
    \acro{nlos}[NLoS]{non-line-of-sight}
    \acro{ap}[AP]{access point}
    \acro{ue}[UE]{user equipment}
    \acro{mimo}[MIMO]{multiple-input multiple-output}
    \acro{mmimo}[mMIMO]{massive multiple-input multiple-output}
    \acro{csi}[CSI]{channel state information}
    \acro{sinr}[SINR]{signal to interference and noise ratio}
    \acro{snr}[SNR]{signal-to-noise ratio}
    \acro{mmse}[MMSE]{minimum mean square error}
    \acro{lmmse}[LMMSE]{local \acl{mmse}}
    \acro{tmmse}[TMMSE]{team \acl{mmse}}
    \acro{ltmmse}[LTMMSE]{local team \acl{mmse}}
    % \acro{mmse}[MMSE]{minimum mean square error}
    % \acro{lmmse}[LMMSE]{local MMSE}
    % \acro{tmmse}[TMMSE]{team minimum mean square error}
    % \acro{ltmmse}[LTMMSE]{local team MMSE}

    \acro{emmse}[E-MMSE]{Element-wise MMSE}
    \acro{mr}[MR]{maximum-ratio}
    \acro{zf}[ZF]{zero-forcing}
    \acro{ls}[LS]{least square}
    \acro{mse}[MSE]{mean squared error}
    \acro{se}[SE]{spectral efficiency}
    \acro{dcc}[DCC]{dynamic cooperation clustering}
    \acro{lsfd}[LSFD]{large scale fading decoding}
    \acro{sota}[SOTA]{state-of-the-art}
    \acro{comp}[CoMP]{coordinated multipoint}
    \acro{cpu}[CPU]{central processing unit}
    \acro{uatf}[UatF]{use-and-then-forget}
    \acro{cd}[CD]{coherent decoding}
\end{acronym}

% paper title
% Titles are generally capitalized except for words such as a, an, and, as,
% at, but, by, for, in, nor, of, on, or, the, to, and up, which are usually
% not capitalized unless they are the first or last word of the title.
% Linebreaks \\ can be used within to get better formatting as desired.
% Do not put math or special symbols in the title.
\title{On the Optimal Performance of Distributed Cell-Free Massive MIMO with LoS Propagation}
%
%
% author names and IEEE memberships
% note positions of commas and nonbreaking spaces ( ~ ) LaTeX will not break
% a structure at a ~ so this keeps an author's name from being broken across
% two lines.
% use \thanks{} to gain access to the first footnote area
% a separate \thanks must be used for each paragraph as LaTeX2e's \thanks
% was not built to handle multiple paragraphs
%

\author{Noor~Ul~Ain,~
        Lorenzo~Miretti,~\IEEEmembership{Member,~IEEE,}
        and~S{\l}awomir~Sta{\'n}czak,~\IEEEmembership{Senior Member,~IEEE}}% <-this % stops a space
        
\author{\IEEEauthorblockN{Noor Ul Ain\IEEEauthorrefmark{1}\thanks{N. U. Ain acknowledge the financial support by the Federal Ministry for Digital and Transport of Germany under Grant 19OI23001A.}, 
 Lorenzo~Miretti\IEEEauthorrefmark{1}\IEEEauthorrefmark{2}\thanks{L. Miretti and S. Sta{\'n}czak acknowledge the financial support by the Federal Ministry of Education and Research of Germany in the programme of “Souver{\"an}. Digital. Vernetzt.” Joint project 6G-RIC, project identification numbers: 16KISK020K and 16KISK030.}
and
S{\l}awomir~Sta{\'n}czak\IEEEauthorrefmark{1}\IEEEauthorrefmark{2}}
\IEEEauthorblockA{\IEEEauthorrefmark{1}Fraunhofer Heinrich Hertz Institute, Berlin, Germany, \{firstname[.middlename].lastname\}@hhi.fraunhofer.de}
\IEEEauthorblockA{\IEEEauthorrefmark{2}Technical University of Berlin, Berlin, Germany}
}

\maketitle
% As a general rule, do not put math, special symbols or citations
% in the abstract or keywords.

\begin{abstract}
In this study, we revisit the performance analysis of distributed beamforming architectures in dense user-centric cell-free \ac{mmimo} systems in \acf{los} scenarios. By incorporating a recently developed optimal distributed beamforming technique, called the \ac{tmmse} technique, we depart from previous studies that rely on suboptimal distributed beamforming approaches for \ac{los} scenarios. Supported by extensive numerical simulations that follow 3GPP guidelines, we show that such suboptimal approaches may often lead to significant underestimation of the capabilities of distributed architectures, particularly in the presence of strong \ac{los} paths. Considering the anticipated ultra-dense nature of cell-free \ac{mmimo} networks and the consequential high likelihood of strong \ac{los} paths, our findings reveal that the team MMSE technique may significantly contribute in narrowing the performance gap between centralized and distributed architectures. %This is motivated by theoretical analysis and supported by extensive numerical simulations that follow 3GPP guidelines. 
\end{abstract}

% Note that keywords are not normally used for peer-reviewed papers.
% \begin{IEEEkeywords}
% IEEE, IEEEtran, journal, \LaTeX, paper, template.
% \end{IEEEkeywords}

% For peer review papers, you can put extra information on the cover
% page as needed:
% \ifCLASSOPTIONpeerreview
% \begin{center} \bfseries EDICS Category: 3-BBND \end{center}
% \fi
%
% For peerreview papers, this IEEEtran command inserts a page break and
% creates the second title. It will be ignored for other modes.
\IEEEpeerreviewmaketitle

\section{Introduction}\label{sec:intro}

\IEEEPARstart{C}{ell-free} \acf{mmimo} has emerged as one of the key research avenues for future generation mobile access networks. Its focus is the study of simple and scalable \ac{ap} cooperation schemes in ultra-dense networks, with the goal of offering uniformly good service to all users \cite{ngo2017cellfree}. Of particular interest is the design and evaluation of efficient transmission techniques for various scalable user-centric cooperation architectures \cite{ngo2017cellfree,9336188,8901451,interdonato2019ubiquitous,atzeni2021distributed,caire2024fairness} under realistic scenarios. The two most common architectures are the (clustered) centralized architecture, which involves sharing both data and \ac{csi} across the \acp{ap} and one or more \acp{cpu}, and the (clustered) distributed architecture, which involves sharing only data.

In the literature, cell-free \ac{mmimo} networks have been optimized and evaluated predominantly by assuming \ac{nlos} propagation models \cite{ngo2017cellfree,9336188,8901451,interdonato2019ubiquitous,atzeni2021distributed,caire2024fairness}. In this case, there are significant performance gaps between distributed and centralized cooperation architectures. However, this assumption overlooks the dense nature of the envisioned cell-free networks, where \ac{los} conditions are more likely. Recently, some studies considered distributed cell-free \ac{mmimo} networks under \ac{los} channel conditions \cite{polegre2020channel,8645336,9740487,8809413,9276421,wang2024optimal}, focusing on aspects such as channel hardening and channel estimation with or without prior knowledge of the phases of the \ac{los} paths. One main limitation of \cite{polegre2020channel,8809413,8645336,9740487,9276421,wang2024optimal} is that they have restricted the analysis to sub-optimal beamforming schemes, hence potentially underestimating the capabilities of distributed architectures.

In contrast to previous works, in this study we consider a recently proposed optimal distributed beamforming scheme in \cite{Miretti2021TeamMP} to reassess the performance of distributed user-centric cell-free \ac{mmimo} networks under strong \ac{los} conditions. More precisely, we consider three beamforming schemes: the optimal centralized scheme (\ac{mmse}) \cite[Eq.~(5.11)]{9336188}; the best known suboptimal distributed scheme (\ac{lmmse}) \cite[Eq.~(5.29)]{9336188}; and the optimal distributed scheme (\ac{ltmmse}) recently derived using the general \ac{tmmse} technique developed in \cite{Miretti2021TeamMP}. Reference \cite{Miretti2021TeamMP} provides a novel beamforming optimization framework that applies to very general channel models and cooperation architectures. As a side observation, \cite{Miretti2021TeamMP} points out that an optimal distributed beamforming design may significantly outperform the \ac{lmmse} scheme under \ac{los} conditions, but leaves a detailed analysis for future work. In this work, we close this gap by providing a comprehensive performance comparison covering important aspects neglected in \cite{Miretti2021TeamMP}, such as: (i) realistic Rician fading model with spatial correlation, phase shifts, and 3GPP-compliant parameters; (ii) channel estimation; (iii) user-centric cooperation clustering; (iv) different \ac{se} bounds; and (v) different power control policies. We also provide an updated discussion on the impact of LoS phase shifts on channel coding and beamforming based on \cite{8809413,9276421,wang2024optimal}, and a missing proof from \cite{Miretti2021TeamMP} on the relation between \ac{lmmse} and \ac{ltmmse} beamforming. Our numerical results demonstrate the \ac{ltmmse} scheme's potential in narrowing the performance gap with the centralized \ac{mmse} scheme in dense networks. 

%A key assumption of our study is the ability of the \acp{ap} to perfectly track the phases of the \ac{los} components. On top of this ideal case, previous works as \cite{8809413} also considered the other extreme case where the phase of the \ac{los} components are completely unknown. As pointed out in \cite{8809413}, the practical system performance fall in between these two scenarios, depending on its phase synchronization/tracking capabilities. Therefore, our performance analysis should be interpreted as an evaluation of the ultimate limits of distributed beamforming architectures. The investigation of more realistic imperfect phase synchronization/tracking regimes is left as an interesting future research direction.

\textit{Paper structure:} Sect. \ref{sec:channel} and Sect. \ref{sec:data_transmission} present the channel and system model, respectively. Sect.~\ref{sec:beamforming} reviews the considered state-of-the-art beamformers in the literature. Our main contribution, i.e., the novel performance comparison, is presented in Sect.~\ref{sec:results}. \textit{Notation:} Lower and upper case bold letters are used for vectors and matrices respectively. The transpose and Hermitian transpose of a matrix $\bm{A}$ are written as $\bm{A}^{\mathsf{T}}$ and $\bm{A}^{\mathsf{H}}$. A block-diagonal matrix with matrices $\bm{D}_1,\dots,\bm{D}_N$ on its diagonal is denoted as $\mathrm{diag}(\bm{D}_1,\dots,\bm{D}_N)$. The expectation of a random variable $X$ is denoted by $\mathbb{E}\{X\}$. %The expectations are taken with respect to all sources of randomness.

\section{Channel and CSI Model}\label{sec:channel}
We consider a user-centric cell-free \ac{mmimo} network with $L$ \acp{ap} indexed by $\mathcal{L} := \{1, \dots, L\}$ jointly serving $K$ \acfp{ue} indexed by $\mathcal{K} := \{1, \dots, K\}$. % that are arbitrarily distributed across a service area. All the \acp{ap} are connected to a \ac{cpu} via a fronthaul network that is error-free \cite{ngo2017cellfree}. 
Each \ac{ap} is equipped with $N$ antennas and each \ac{ue} is equipped with a single antenna. For simplicity, we focus on the \ac{ul},  but we remark that our conclusions can be rather straightforwardly 
adapted to the \ac{dl} in time-division duplex systems by using known channel reciprocity \cite{ngo2017cellfree,9336188} and UL-DL duality arguments \cite{9336188,miretti2023uldl}. 

\subsection{Spatially Correlated Rician Fading with Phase Shifts}
To cover \ac{los} propagation, we consider a spatially correlated Rician fading channel model with phase shifts. Following the block fading assumption, the channel remains time-invariant and frequency-flat within a coherence block of $\tau_c$ symbols and evolves across coherence blocks according to a stationary and ergodic random process. In an arbitrary coherence block, we let $\bm{h}_{k,l}\in\mathbb{C}^N$ be a realization of the channel coefficients between \ac{ue} $k\in\mathcal{K}$ and the $N$ antennas of \ac{ap} $l\in\mathcal{L}$. We assume $(\forall k\in\mathcal{K})(\forall l\in\mathcal{L})$
\begin{equation}\label{eq:rician}
\bm{h}_{k,l} = \bm{\bar{h}}_{k,l}e^{j\theta_{k,l}}+\tilde{\bm{h}}_{k,l}, \quad \tilde{\bm{h}}_{k,l}\sim\mathcal{N}_{\mathbb{C}}\left(\bm{0},\bm{R}_{k,l}\right)
\end{equation}
where $\bm{\bar{h}}_{k,l}\in\mathbb{C}^N$ is the spatial signature of the deterministic \ac{los} component, $\theta_{k,l}\in [0,2\pi]$ is the associated phase shift, and $\bm{R}_{k,l}\in\mathbb{C}^{N\times N}$ is a covariance matrix describing the spatial correlation induced by the \ac{nlos} components.

Departing from the canonical Rician fading model \cite{polegre2020channel,8645336,9740487}, the previous works \cite{8809413,9276421} let the LoS phases be random variables $\theta_{k,l}\sim \mathcal{U}[0,2\pi]$ that evolve across coherence blocks as rapidly as $\tilde{\bm{h}}_{k,l}$, and study the system performance by assuming either perfectly known or completely unknown LoS phases. This was motivated in \cite{8809413} by arguing that $\theta_{k,l}$ varies much faster than $\bm{\bar{h}}_{k,l}$ and $\bm{R}_{k,l}$, and hence it may not be easily estimated, particularly in very high mobility scenarios and for single-antenna APs. However, as also pointed out in \cite{8809413,9276421}, the practical system performance fall in between these two scenarios. In addition, in practice, $\theta_{k,l}$ does not vary as rapidly as $\tilde{\bm{h}}_{k,l}$, whose variations are dominated by the delay and Doppler \textit{spread} of the entire multipath channel. For instance, as also observed by the authors of \cite{8809413,9276421} in the recent work \cite{wang2024optimal}, $\theta_{k,l}$ can be safely assumed constant over many coherence blocks in the frequency domain.

In this work, we assume that the network can perfectly track the \ac{los} phases, similar to \cite{wang2024optimal} and to the \textit{phase-aware} scenario in \cite{8809413,9276421}. Moreover, we assume that each realization of $\theta_{k,l}\sim \mathcal{U}[0,2\pi]$ is kept constant over multiple coherence blocks, i.e., realizations of $\tilde{\bm{h}}_{k,l}$, as in \cite{wang2024optimal}, and hence potentially treated as a large-scale fading parameter (statistical \ac{csi}) from a channel coding and beamforming optimization perspective (see Remark~\ref{rem:ergodic} and Remark~\ref{rem:conditioning}). Differently than \cite{8809413,9276421}, we will not focus on the impact of different LoS phase tracking capabilities, but rather on the impact of different beamforming architectures under perfect LoS phase tracking. The extension to imperfect LoS phase tracking is left as an interesting future direction. 

%The channel vector $\bm{h}_{k,l}$ is modeled as an independent circularly-symmetric complex Gaussian random vector $(\forall k\in\mathcal{K})(\forall l\in\mathcal{L})$
%\begin{equation}\label{eq:channel}
%\bm{h}_{k,l}\sim\mathcal{N}_{\mathbb{C}}\left(\bm{\bar{h}}_{k,l},\bm{R}_{k,l}\right), 
%\end{equation}
%where the expected value $\bm{\bar{h}}_{k,l}\in\mathbb{C}^N$ represents the \ac{los} component and $\bm{R}_{k,l}\in\mathbb{C}^{N\times N}$ is a covariance matrix describing the spatial correlation induced by the \ac{nlos} components.

\subsection{Uplink Channel Estimation}\label{sec:CSI}
We assume that the network uses \ac{ul} pilot transmission followed by (phase-aware) \ac{mmse} channel estimation \cite{8809413,9276421,wang2024optimal} to acquire \ac{csi}. For pilot signaling, the network employs $\tau_p$ out of $\tau_c$ symbols in a coherence block to transmit $\tau_p$ mutually orthogonal pilot sequences. We assume that the number of \acp{ue} is large such that $\tau_p\ll K$, thus some pilots must be shared by more than one \ac{ue}. We denote the set of \acp{ue} sharing the pilot with \ac{ue} $k\in\mathcal{K}$ as $\mathcal{P}_k$. In the \ac{ul} pilot transmission phase, the received signal at \ac{ap} $l\in\mathcal{L}$, after decorrelating with respect to the pilot $t_k\in \{1,\ldots,\tau_p\}$ of user $k\in \mathcal{K}$, is given by \cite{9336188}
\begin{equation*}
\bm{y}_{t_k,l}^{\text{pilot}}=\sqrt{\eta_k}\tau_p\bm{h}_{k,l}+\sum_{i\in\mathcal{P}_k/\{k\}}\sqrt{\eta_i}\tau_p\bm{h}_{i,l}+\bm{n}_{t_k,l}^{\text{pilot}},
\end{equation*}
where $\eta_i\in\mathbb{R}_+$ is the pilot transmit power of \ac{ue} $i\in\mathcal{K}$,  the first term represents the desired signal from \ac{ue} $k$, the middle term represents the pilot contamination effect, and $\bm{n}_{t_k,l}^{\text{pilot}}\sim \mathcal{N}_\mathbb{C}(\bm{0},\sigma_{\text{ul}}^2\tau_p\bm{I}_N)$ is the receiver noise.

The (phase-aware) \ac{mmse} estimate $\bm{\hat{h}}_{k,l}$ of $\bm{h}_{k,l}$ from $\bm{y}_{t_k,l}^{\text{pilot}}$ is given by \cite{8809413,9276421} $(\forall k\in\mathcal{K})(\forall l\in\mathcal{L})$
\begin{equation*}
\bm{\hat{h}}_{k,l}=\bm{\bar{h}}_{k,l}e^{j\theta_{k,l}}+\sqrt{\eta_k}\bm{R}_{k,l}\bm{\Psi}_{t_k,l}^{-1}\left(\bm{y}_{t_k,l}^{\text{pilot}}-\bm{\bar{y}}_{t_k,l}\right),
\end{equation*}
where $\bm{\bar{y}}_{t_k,l}=\sum_{i\in\mathcal{P}_k}\sqrt{\eta_i}\tau_p\bm{\bar{h}}_{i,l}e^{j\theta_{i,l}}$, and $\bm{\Psi}_{t_k,l}=\sum_{i\in\mathcal{P}_k}\eta_i\tau_p\bm{R}_{i,l}+\sigma_{\text{ul}}^2\bm{I}_N.$ Note that in the above, on top of perfect knowledge of the LoS phase, we also assume that all statistical parameters, such as $\bm{\bar{h}}_{k,l}$, $\bm{R}_{k,l}$, and $\sigma_{\text{ul}}^2$, are known by the network, as customary in the literature. The estimation error $\bm{\xi}_{k,l}=\bm{{h}}_{k,l}-\bm{\hat{h}}_{k,l}$ has zero mean and covariance matrix $
\bm{C}_{k,l} = \bm{R}_{k,l} − \eta_k\tau_p\bm{R}_{k,l}\bm{\Psi}_{t_k,l}^{-1}\bm{R}_{k,l}$. Furthermore, for fixed $\theta_{k,l}$, the local estimate $\bm{\hat{h}}_{k,l}$ and the estimation error $\bm{\xi}_{k,l}$ are independent random vectors distributed as $\mathcal{N}_{\mathbb{C}}(\bm{\bar{h}}_{k,l},\bm{R}_{k,l}-\bm{C}_{k,l})$ and $\mathcal{N}_{\mathbb{C}}(\bm{0},\bm{C}_{k,l})$, respectively.

 %Although important, an analysis on the imperfect knowledge of these quantities is out of the scope of this work, since we focus on the fundamental limits of cell-free \ac{mmimo}. 
For convenience, we denote by 
$\bm{\hat{H}}_{l}=[\bm{\hat{h}}_{1,l},\dots, \bm{\hat{h}}_{K,l}]\in\mathbb{C}^{N\times K}$ the local  estimate of the local channel $\bm{{H}}_{l}=[\bm{{h}}_{1,l},\dots, \bm{{h}}_{K,l}]\in\mathbb{C}^{N\times K}$ of the $l$th \ac{ap}, by $\bm{\hat{H}}=[\bm{\hat{H}}_{1}^{\mathsf{T}},\dots,\bm{\hat{H}}_{L}^{\mathsf{T}}]^{\mathsf{T}}\in\mathbb{C}^{LN\times K}$ the global estimate of the global channel $\bm{{H}}=[\bm{{H}}_{1}^{\mathsf{T}},\dots,\bm{{H}}_{L}^{\mathsf{T}}]^{\mathsf{T}}\in\mathbb{C}^{LN\times K}$, and by $\bm{\hat{h}}_{k}=[\bm{\hat{h}}_{k,1}^{\mathsf{T}},\dots, \bm{\hat{h}}_{k,L}^{\mathsf{T}}]^{\mathsf{T}}\in\mathbb{C}^{LN}$ the global estimate of the concatenated channel $\bm{{h}}_{k}=[\bm{{h}}_{k,1}^{\mathsf{T}},\dots, \bm{{h}}_{k,L}^{\mathsf{T}}]^{\mathsf{T}}\in\mathbb{C}^{LN}$ of the $k$th \ac{ue}.

\section{Uplink Data Transmission}\label{sec:data_transmission}
In an arbitrary time-frequency resource element, the \ac{ul} data signal $\bm{y}_l^{\text{ul}}\in\mathbb{C}^N$ received at \ac{ap} $l$  is given by $(\forall l\in\mathcal{L})$
\begin{equation*}
\bm{y}_l^{\text{ul}}=\sum_{k=1}^{K}\sqrt{p_k}\bm{h}_{k,l}s_{k}+\bm{n}_l^{\text{ul}},
\end{equation*}
where $s_{k}\sim\mathcal{N}_{\mathbb{C}}(0,1) $ is the data signal sent from \ac{ue} $k\in\mathcal{K}$ with power $p_{k}\geq 0$, and $\bm{n}_l^{\text{ul}}\sim\mathcal{N}_{\mathbb{C}}(\bm{0}, \sigma^2_{\text{ul}}\bm{I}_N)$ is the additive noise at \ac{ap} $l$. The received signals $\bm{y}^{\text{ul}}=\big[\bm{y}_{1}^{\text{ul}^\mathsf{T}},\dots,\bm{y}_{L}^{\text{ul}^\mathsf{T}}\big]^{\mathsf{T}}$ from all the \acp{ap} are then combined using network-wide beamforming vectors $\bm{v}_{k}=\big[\bm{v}^{\mathsf{T}}_{k,1},\dots,\bm{v}^{\mathsf{T}}_{k,L}\big]^{\mathsf{T}}$. % to separate the desired data $s_k$ from the rest of the interference in the network as accurately as possible. 
To incorporate an arbitrary \ac{dcc} scheme \cite{9336188}, we let $\mathcal{L}_k$ be the subset of \acp{ap} serving \ac{ue} $k\in\mathcal{K}$, and define the matrices $(\forall k\in\mathcal{K})$ $\bm{D}_k= \mathrm{diag}(\bm{D}_{k,1},\dots,\bm{D}_{k,L})$, where $(\forall l\in\mathcal{L})$ $
\bm{D}_{k,l}= \bm{I}_N$ if $l\in \mathcal{L}_k$, and $\bm{D}_{k,l}=\bm{0}_N$ otherwise. The estimate of $s_k$ is then given by 
\begin{equation}\label{eq:data_detection}
(\forall k\in\mathcal{K})~\hat{s}_{k}=\sum_{l=1}^L{\bm{v}_{k,l}^{\mathsf{H}}\bm{D}_{k,l}\bm{y}_l^{\text{ul}}}=\bm{v}_{k}^{\mathsf{H}}\bm{D}_{k}\bm{y}^{\text{ul}}.
\end{equation}

We evaluate the performance of the cell-free \ac{mmimo} network using two performance metrics. The first metric is a state-of-the-art lower bound on the ergodic \ac{se}, known as the \textit{\ac{uatf}} bound, which is given by \cite{9336188}  
\begin{equation}\label{eq:se_uatf}
(\forall k\in\mathcal{K})~\text{SE}^{\text{ul,UatF}}_k = \frac{\tau_c-\tau_p}{\tau_c}\log_2\left(1+\text{SINR}^{\text{ul,UatF}}_k\right),
\end{equation}
where the effective SINR is given by $\text{SINR}^{\text{ul,UatF}}_k=$
\begin{equation*}
\frac{p_k\left|\mathbb{E}\left\{g_{kk}\right\}\right|^2}{\sum_{i=1}^K p_i\mathbb{E}\left\{\left|g_{ki}\right|^2\right\}-p_k\left|\mathbb{E}\left\{g_{kk}\right\}\right|^2+\sigma_{\text{ul}}^2\mathbb{E}\left\{\| \bm{D}_k\bm{v}_k\|^2\right\}},
\end{equation*}
where $g_{ki} = \bm{v}^{\mathsf{H}}_k\bm{D}_k\bm{h}_i$. This bound is derived by assuming that the available CSI is first exploited for receive beamforming, and then discarded in the channel decoding phase. The second metric is a more conventional lower bound obtained under the assumption that \ac{csi} is available at the decoder. It is known as the \acf{cd} bound, and it is given by \cite{9336188} 
\begin{equation}\label{eq:se_cd}
(\forall k\in\mathcal{K})~\text{SE}^{\text{ul,cd}}_k = \frac{\tau_c-\tau_p}{\tau_c}\mathbb{E}\left\{\log_2\left(1+\text{SINR}^{\text{ul,cd}}_k\right)\right\},
\end{equation}
where the instantaneous SINR is given by $\text{SINR}^{\text{ul,cd}}_k=$ 
\begin{equation*}
\frac{p_k|\bm{v}^{\mathsf{H}}_k\bm{D}_k\bm{\hat{h}}_k|^2}{\sum_{\substack{i=1 \\ i\neq j}}^K p_i|{\bm{v}_{k}^{\mathsf{H}}\bm{D}_{k}\bm{\hat{h}}_i|^2+\bm{v}_{k}^{\mathsf{H}}\bm{D}_{k}\bm{Z}\bm{D}_{k}\bm{v}_k+\sigma_{\text{ul}}^2\|\bm{D}_k\bm{v}_k\|^2}},\\
\end{equation*}
where $\bm{Z}=\sum_{i=1}^Kp_i\bm{C}_{i}$, $\bm{C}_{k}=\mathrm{diag}(\bm{C}_{k,1},\dots,\bm{C}_{k,L})$. We remark that \eqref{eq:se_cd} offers a more realistic lower-bound on the achievable \ac{se}, but, due to the expectation operation in front of the logarithm, it often leads to intractable optimization problems, especially in distributed setups \cite{miretti2023uldl}.
\begin{remark}\label{rem:ergodic}
In this work, the expectations in \eqref{eq:se_uatf} and \eqref{eq:se_cd} are evaluated for fixed LoS phases. Operationally, this can be interpreted as coding over many coherence blocks with constant LoS phases, which is consistent with the channel model in Sect.~\ref{sec:channel}. However, the results in this work could be readily extended to much longer codewords spanning many realizations of the LoS phases by taking the expectations in \eqref{eq:se_uatf} and \eqref{eq:se_cd} with respect to all sources of randomness.  
\end{remark}

\section{Beamforming Schemes}\label{sec:beamforming}
In this section we review and connect state-of-the-art centralized and distributed beamforming schemes by focusing on a generalized \ac{mse} criterion, i.e., by considering the following optimization problem \cite{Miretti2021TeamMP}: $(\forall k\in\mathcal{K})$
\begin{equation}\label{eq:MSE}
\minimize_{\bm{v}_k\in\mathcal{V}_k}\quad \mathbb{E}\left\{\left|s_k-\hat{s}_k\right|^2\right\},
\end{equation}
where $\hat{s}_k$ is given by \eqref{eq:data_detection}, and where $\mathcal{V}_k$ denotes a given set of functions mapping the available \ac{csi} to beamforming coefficients in $\mathbb{C}^{LN}$. More precisely, we let $\mathcal{V}_k$ be a given \textit{subspace} of the space of functions mapping realizations of $\bm{\hat{H}}$ to realizations of $\bm{v}_k$. We refer to \cite{Miretti2021TeamMP} for additional mathematical details on the definition of these constraints. As already discussed, two main beamforming implementations are commonly considered in the literature, typically referred to as the centralized and distributed schemes. These schemes are characterized by varying levels of cooperation between the \acp{ap}, in particular with respect to the level of \ac{csi} sharing\footnote{Some works such as \cite{Miretti2021TeamMP} use the term \textit{distributed} beamforming to denote general beamforming architectures with arbitrary levels of CSI sharing, hence including centralized beamforming as a particular case. In this work, we follow the terminology in \cite{9336188}, where distributed beamforming refers to the particular case of no instantaneous CSI sharing.}. As done in \cite{Miretti2021TeamMP,miretti2023uldl}, different levels of \ac{csi} sharing can be formally included in \eqref{eq:MSE} using the constraint $\mathcal{V}_k$. In the following, we will informally review how to map the considered beamforming implementations to appropriate $\mathcal{V}_k$, and how to produce optimal solutions to \eqref{eq:MSE}. Importantly, we remark that the solution to \eqref{eq:MSE} not only maximizes \eqref{eq:se_cd} for centralized beamforming architectures \cite{9336188}, but also generally maximizes the \ac{uatf} bound on ergodic \ac{ul} rates given in \eqref{eq:se_uatf} under general beamforming architectures \cite{Miretti2021TeamMP,miretti2023uldl}. This second observation is often overlooked in the literature.

\begin{remark}\label{rem:conditioning}
In analogy with Remark~\ref{rem:ergodic}, all expectations in this section are evaluated for fixed LoS phases, which we recall are assumed perfectly known by the network. Equivalent expressions for the case of random (yet perfectly known) phases can be readily obtained by replacing all expectations with conditional expectations given the LoS phases.
\end{remark}

\subsection{Centralized Beamforming}
In a centralized cell-free network, the data detection is carried out under the assumption that imperfect \ac{csi} is perfectly shared within the serving cluster of each \ac{ue} $k\in\mathcal{K}$. Following \cite{Miretti2021TeamMP,miretti2023uldl}, this can be modeled by letting $\mathcal{V}_k$ in \eqref{eq:MSE} be the full space of functions of the global \ac{csi} $\hat{\bm{H}}$. In this case, the optimal solution to \eqref{eq:MSE} is derived by decomposing the problem into disjoint conditional \ac{mmse} problem, one for each realization of $\hat{\bm{H}}$, expressed as $(\forall k\in\mathcal{K})$
\begin{equation*}
\minimize_{\bm{v}_{k}\in\mathbb{C}^{LN}}\;\mathbb{E}\left\{\left|s_k-\bm{v}_{k}^{\mathsf{H}}\bm{D}_{k}\bm{y}^{\text{ul}}\right|^2\:\big|\:\bm{\hat{H}}\right\}.
\end{equation*}
The resulting optimal beamforming vector takes the form of the well-known centralized MMSE solution \cite[Eq.~(5.11)]{9336188}. With our notation\footnote{The main difference with respect to \cite{9336188} is that we consider $s_k\sim \mathcal{N}_{\mathbb{C}}(0, 1)$ instead of $s_k\sim \mathcal{N}_{\mathbb{C}}(0, p_k)$. The two models are completely equivalent, in the sense that they lead to identical transmit signals and achievable rates, although the MSE-optimal beamformers differ by a scaling factor.}, it is given by $(\forall k\in\mathcal{K})~\bm{v}_k^{\text{MMSE}}=$
% \begin{equation}\label{eq:mmse}
% \sqrt{p_k}\left(\sum^K_{i=1}{p_i\bm{D}_k\left(\bm{\hat{h}}_i\bm{\hat{h}}_i^{\mathsf{H}}+\bm{C}_i\right)\bm{D}_k}+\sigma_{\text{ul}}^2\bm{I}_{LN}\right)^{-1}\bm{D}_k\bm{\hat{h}}_k.
% \end{equation}
\begin{equation}\label{eq:mmse}(\bm{D}_k\hat{\bm{H}}\bm{P}\hat{\bm{H}}^{\mathsf{H}}\bm{D}_k+\bm{D}_{k}\bm{Z}\bm{D}_{k}+\sigma_{\text{ul}}^2\bm{I}_{LN})^{-1}\bm{D}_k\hat{\bm{H}}\bm{P}^{\frac{1}{2}}\bm{e}_k,
\end{equation}
where $\bm{P}= \mathrm{diag}(p_1,...,p_K)$, and $\bm{e}_k$ is the $k$th column of $\bm{I}_{K}$. Note that $\bm{D}_k\hat{\bm{H}}$ can be computed using only the channels of the APs belonging to the cluster $\mathcal{L}_k$ of UE $k$. It is well-known that the vector \eqref{eq:mmse} that minimizes the \ac{mse} in data detection also maximizes the coherent decoding lower bound \eqref{eq:se_cd} on the \ac{ul} ergodic rates \cite{9336188}. A less known fact is that it also maximizes the UatF bound in \eqref{eq:se_uatf} \cite{Miretti2021TeamMP,miretti2023uldl}.

\subsection{Distributed Beamforming}
In a distributed cell-free network, there is no instantaneous \ac{csi} sharing within each cluster (only the slowly-varying statistical CSI and LoS phases are shared, as already discussed in Sect.~\ref{sec:channel}). Each \ac{ap} $l\in\mathcal{L}_k$ performs beamforming locally based on local CSI, to obtain local data estimates $\hat{s}_{k,l}$. The local estimates from all the serving \acp{ap} are then combined at the decoder \cite{9336188}. Following \cite{Miretti2021TeamMP,miretti2023uldl}, this constraint can be modelled mathematically by letting $\mathcal{V}_k$ in \eqref{eq:MSE} be the subspace of (vector-valued) functions of $\hat{\bm{H}}$ where each $N$-dimensional subvector depends only on the local \ac{csi} $\hat{\bm{H}}_l$ of the corresponding \ac{ap} (and fixed problem parameters such as the statistical CSI and the LoS phases). Due to this non-trivial constraint, \eqref{eq:MSE} cannot be solved by decomposing it into disjoint conditional MMSE problems as for the centralized case.

\subsubsection{Local MMSE with Optimal LSFD}
To circumvent this issue, a suboptimal distributed beamforming scheme, known as \acf{lmmse} beamforming, attempts to calculate the beamforming vectors by optimizing each local \ac{mse} $\mathbb{E}\{\left|s_k-\hat{s}_{k,l}\right|^2\}$ separately for each AP. Specifically, in analogy with the centralized \ac{mmse} beamforming, it achieves a suboptimal solution to \eqref{eq:MSE} by solving $(\forall k\in\mathcal{K})(\forall l\in\mathcal{L})$
\begin{equation}\label{eq:mse_dist}
\minimize_{\bm{v}_{k,l}\in\mathbb{C}^{N}}\;\mathbb{E}\left\{\left|s_k-\bm{v}_{k,l}^{\mathsf{H}}\bm{D}_{k,l}\bm{y}_l^{\text{ul}}\right|^2\:\Big|\:\bm{\hat{H}}_l\right\}.
\end{equation}
The resulting beamforming vector $\bm{v}_{k,l}$ takes the form of \cite[Eq.~(5.29)]{9336188}. With our notation, it is given by the $k$th column of $(\forall k\in\mathcal{K})(\forall l\in\mathcal{L})$
\begin{equation}\label{eq:lmmse_T}
\bm{V}_{l}=(\hat{\bm{H}}_l\bm{P}\hat{\bm{H}}_l^{\mathsf{H}}+\bm{Z}_l+\sigma_{\text{ul}}^2\bm{I}_{N})^{-1}\hat{\bm{H}}_l\bm{P}^{\frac{1}{2}},
\end{equation}
where $\bm{Z}_l = \sum_{i=1}^Kp_i\bm{C}_{i,l}$. Subsequently, each local beamforming vector $\bm{v}_{k,l}$ is assigned a correcting weight $c_{k,l}\in\mathbb{C}\,(\forall k\in\mathcal{K}) (\forall l\in\mathcal{L})$. These correcting weights, determined at the decoder using only the statistical \ac{csi} (and knowledge of the LoS phases, in our setup), are called \textit{\acf{lsfd}} weights \cite{9336188}. The final \ac{lmmse} with optimal \ac{lsfd} beamforming vector  becomes $(\forall k \in \mathcal{K})(\forall l \in \mathcal{L}_k)$
 \begin{equation}\label{eq:lmmse_lsfd}
\bm{v}_{k,l}^{\text{LMMSE - lsfd}}=\bm{V}_{l}\bm{e}_kc_{k,l}^{\text{lsfd}},
\end{equation}
where $c_{k,l}^{\text{lsfd}}$ are chosen to maximize the \ac{sinr} of \ac{uatf} bound in \eqref{eq:se_uatf} as in \cite[Eq.~(5.30)]{9336188} (maximization of Rayleigh quotient). It can be shown that this is also equivalent to minimizing the MSE $\mathbb{E}\{\left|s_k-\hat{s}_k\right|^2\}$ with respect to the LSFD weights.

The suboptimal approach described above optimizes the beamformers of each \ac{ap} disjointly, by neglecting the impact of the other \acp{ap}, except for the optimization of the LSFD weights. Consequently, the derived \ac{lmmse} beamforming vectors do not necessarily achieve the network-wide optimality, i.e., the optimum of \eqref{eq:MSE} under the given constraint $\mathcal{V}_k$ modeling no \ac{csi} sharing. Despite the improved coordination offered by the design of the \ac{lsfd} weights, a more sophisticated approach is necessary to overcome this limitation.

\subsubsection{Local Team MMSE}
The recently proposed \ac{ltmmse} beamforming technique \cite{Miretti2021TeamMP} introduces an optimal solution method for Problem \eqref{eq:MSE}, and hence for maximizing the UatF bound in \eqref{eq:se_uatf}, under general beamforming architectures (i.e., constraint $\mathcal{V}_k$). Once specialized to the case of distributed beamforming with no instantaneous CSI sharing, it is based on the fact that the optimal beamforming vector $\bm{v}_{k,l}$ for \ac{ap} $l\in\mathcal{L}$ must satisfy the necessary optimality conditions given by the solution to the following optimization problem $(\forall k\in\mathcal{K})(\forall l\in\mathcal{L})$
\begin{equation*}
\minimize_{\bm{v}_{k,l}\in\mathbb{C}^{N}}\;\mathbb{E}\Big\{\big|s_k-\bm{v}_{k,l}^{\mathsf{H}}\bm{D}_{k,l}\bm{y}^{\text{ul}}_l - \sum_{j \in \mathcal{L} / l}\underbrace{\bm{v}_{k,j}^{\mathsf{H}}}_{\text{fixed}}\bm{D}_{k,j}\bm{y}^{\text{ul}}_j\big|^2\:\Big|\:\bm{\hat{H}}_l\Big\}.
\end{equation*}

From a mathematical point of view, these conditions are reminiscent of the game theoretical notion of Nash equilibrium, although, strictly speaking, we are not in a game theoretical setting since \eqref{eq:MSE} is a single objective optimization problem. The team theoretical framework in \cite{Miretti2021TeamMP} proves that these conditions are not only necessary but also sufficient for optimality. Thus, an optimal solution can be found by solving this set of optimality conditions jointly across all the \acp{ap}. The resulting optimal \ac{ltmmse} beamforming vector is given by \cite[Thm.~4]{Miretti2021TeamMP}, \cite{miretti2023uldl}[Prop.~11] $(\forall k\in\mathcal{K})(\forall l\in\mathcal{L})$
\begin{equation}\label{eq:tmmse}
\bm{v}_{k,l}^{\text{LTMMSE}} = \bm{V}_{l}\bm{c}_{k,l} \quad \forall l \in \mathcal{L}_k,   %Nx1 
\end{equation}
with $\bm{V}_{l}$ taking the same form as in the \ac{lmmse} beamforming matrix \eqref{eq:lmmse_T}, which is computed using the \ac{ap}'s local \ac{csi}, and where the vector $\bm{c}_{k,l} \in \mathbb{C}^K$ denotes a second decoding stage, which is computed by the cluster processor using channel statistics and knowledge of the LoS phases. %applying a further correction to the local beamforming vector by taking into account the effects of the other \acp{ap}.
The optimal $\bm{c}_{k,l}$ is calculated by letting $(\forall l\in\mathcal{L})~\bm{\Pi}_l=\mathbb{E}\{\bm{P}^{\frac{1}{2}}\hat{\bm{H}}_l^{\mathsf{H}}\bm{V}_l\}$ and by solving the system of linear equations \cite[Thm.~4]{Miretti2021TeamMP}, \cite{miretti2023uldl}[Prop.~11] $(\forall k\in\mathcal{K})$
\begin{equation}\label{eq:lin_syst}
\begin{cases}
    \bm{c}_{k,l} + \sum_{j \in \mathcal{L}_k / l} \bm{\Pi}_j \bm{c}_{k,j} = \bm{e}_k & \forall l \in\mathcal{L}_k,\\
    \bm{c}_{k,l}=\bm{0}_{K\times1} & \text{otherwise,}
    \end{cases}
\end{equation}
 % This second decoding vector $\bm{c}_{k,l}$ optimally enhances \ac{lmmse} beamforming by taking into account the long-term impact of the other \acp{ap}. 

\subsection{Impact of LoS propagation}
The next proposition shows that, in the case of fully \ac{nlos} propagation and no pilot contamination, the \ac{ltmmse} beamforming vector $\bm{v}_{k}^{\text{LTMMSE}}$ boils down to the \ac{lmmse} beamforming vector with optimal \ac{lsfd} weights as in \eqref{eq:lmmse_lsfd}. This was already observed in \cite{Miretti2021TeamMP} without proof.
\begin{proposition}\label{prop:equivalence}
If $\hat{\bm{h}}_{k,l}$ is independently distributed as $\mathcal{N}_{\mathbb{C}}(\bm{0},\bm{R}_{k,l}-\bm{C}_{k,l})$ for all $k\in \mathcal{K}$ and $l\in \mathcal{L}$, then \begin{equation*}
(\forall k\in \mathcal{K})(\forall l\in \mathcal{L})~\bm{v}_{k,l}^{\text{LMMSE - lsfd}} = \bm{v}_{k,l}^{\text{LTMMSE}}.
\end{equation*}
\begin{proof}
The $(i,j)$th entry of $\bm{\Pi}_l$ can be written as $[\bm{\Pi}_l]_{i,j}=\mathbb{E}\{\sqrt{p_i}\hat{\bm{h}}_{i,l}^{\mathsf{H}}(\sum_{k\in \mathcal{K}}p_k\hat{\bm{h}}_{k,l}\hat{\bm{h}}_{k,l}^{\mathsf{H}}+\bm{Z}_l+\sigma_{\text{ul}}^2\bm{I}_{N})^{-1} \hat{\bm{h}}_{j,l}\sqrt{p_j}\}$. Since $\hat{\bm{h}}_{i,l}~\sim -\hat{\bm{h}}_{i,l}$, and since $\hat{\bm{h}}_{i,l}$ is independent of everything else, we observe that $[\bm{\Pi}_l]_{i,j} = -[\bm{\Pi}_l]_{i,j}$ for $i\neq j$, which implies $[\bm{\Pi}_l]_{i,j}=0$ for $i\neq j$. Then, since all $\bm{\Pi}_l$ are diagonal, the optimal vectors $\bm{c}_{k,l}$ solving \eqref{eq:lin_syst} boil down to $c_{k,l}\bm{e}_k$.
\end{proof}
\end{proposition}
However, in the case of \ac{los} propagation, \ac{ltmmse} beamforming may give larger \ac{se} compared to \ac{lmmse} beamforming. This discrepancy becomes evident in the extreme case where the \ac{los} component is dominant. In this case, \ac{ltmmse} beamforming approaches \ac{mmse} beamforming \cite{Miretti2021TeamMP}. This is formalized in the next proposition.  
\begin{proposition}\label{prop:equivalence2}
If $\hat{\bm{h}}_{k,l} = \bm{\bar{h}}_{k,l}e^{j\theta_{k,l}}$ for all $k\in \mathcal{K}$ and $l\in \mathcal{L}$, then the $l$th subvector of $\bm{v}_{k}^{\text{MMSE}}$ satisfies
\begin{equation*}
(\forall k\in \mathcal{K})(\forall l\in \mathcal{L})~\bm{v}_{l,k}^{\text{MMSE}} = \bm{v}_{k,l}^{\text{LTMMSE}}.
\end{equation*}
\begin{proof}
We observe that the centralized solution \eqref{eq:mmse} is also feasible in the distributed case, since $\hat{\bm{h}}_{k,l} =\bm{\bar{h}}_{k,l}e^{j\theta_{k,l}}$ is a fixed parameter known by the network. Since the space of feasible distributed beamformers is a subspace of the space of centralized beamformers, and since \eqref{eq:MSE} has a unique solution in both cases \cite{Miretti2021TeamMP}, the two solutions must coincide.   
\end{proof}
\end{proposition}
We point out the importance of the second beamforming stage since $\bm{c}_{k,l}$ in \eqref{eq:tmmse}, which gives enough flexibility to implement the \ac{mmse} solution \eqref{eq:mmse} with $\hat{\bm{H}}$ replaced by its mean. In contrast, the single \ac{lsfd} coefficient in \eqref{eq:lmmse_lsfd} does not give enough flexibility.

\section{Numerical Results}\label{sec:results}

\subsection{Parameters and Setup}

\begin{table}[h!]
\centering
\renewcommand{\arraystretch}{1.2} % Increases row spacing
\setlength{\tabcolsep}{10pt} % Adjusts the space between columns
\begin{tabular}{||c c||} 
\hline
\textbf{Parameter} & \textbf{Value} \\ [1ex]
\hline\hline
Network area & $d$ × $d$, $d\in [200,1000]$ m \\
\hline
Network layout & Random deployment \\
\hline
Number of \acp{ap} & $L=100$\\
\hline
Number of UEs & $K=40$\\
\hline
Number of antennas per AP & $N=4$ \\
\hline
Bandwidth & $B=100$ MHz\\
\hline
Carrier frequency & $f_c=5$ GHz\\
\hline
Maximum \ac{ul} transmit power & $p_{\text{max}}\in [20,100]$ mW\\
\hline
Coherence block symbols & $\tau_c=200$\\
\hline
Pilot symbols & $\tau_p=5$\\
\hline
AP-UE height difference & $\Delta h=11$ m\\
\hline
Shadow fading \ac{los} & $\sigma_{\text{sf}} = 8$ dB \\
\hline
Antenna spacing & $d=\lambda/2$\\
\hline
\end{tabular}\\
\vspace{3mm} % Add space between the table and the caption
\caption{Simulation parameters}
\label{tab:parameters}
\end{table}

\begin{figure*}[t!]
\centering
\subfloat[]
{\includegraphics[width=0.33\textwidth]{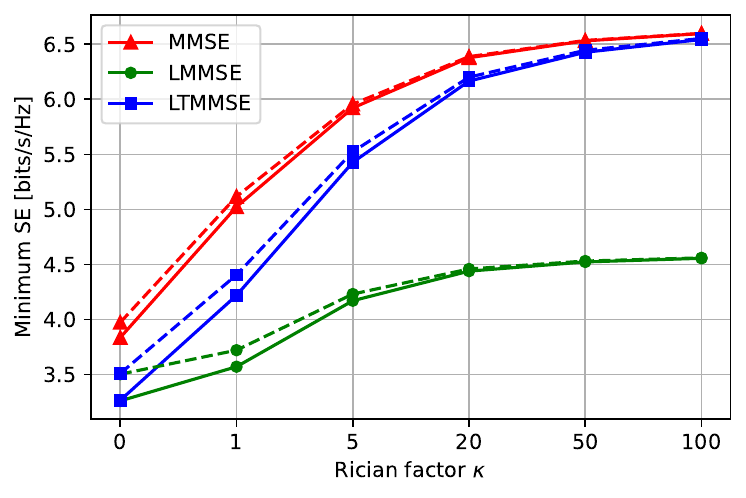}\label{fig:result_k_minse}}
\hfil
\subfloat[]
{\includegraphics[width=0.33\textwidth]{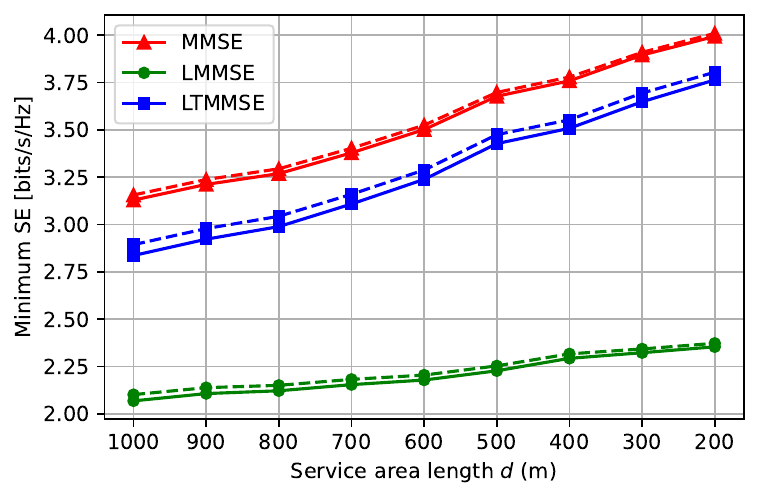}\label{fig:result_d_minse}}
\hfil
\subfloat[]
{\includegraphics[width=0.33\textwidth]{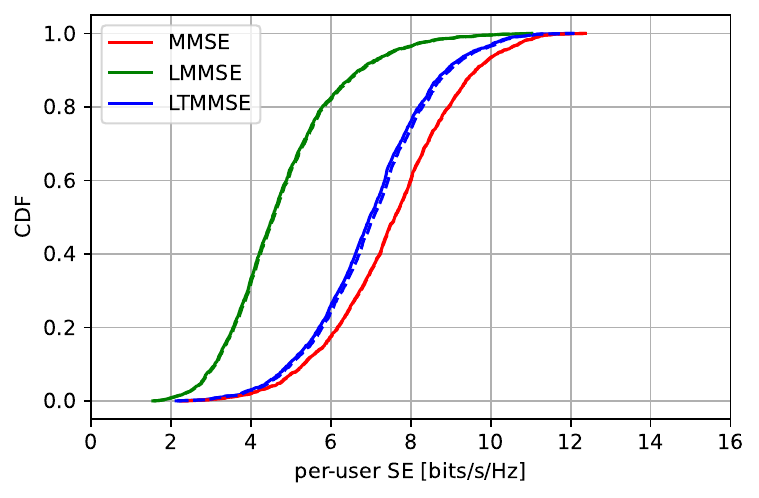}\label{fig:result_cdf_minse}}
\caption{Comparison of \ac{ul} SEs achieved by different beamforming schemes for the case $v=-1$ in \eqref{eq:power_control}. The solid lines refer to the \ac{uatf} bound \eqref{eq:se_uatf}, and the dotted lines refer to the coherent decoding bound \eqref{eq:se_cd}. a)  Minimum \ac{ul} \ac{se} for different values of $\kappa$ ($d=1$ km, $p_{\max} = 100$ mW); b) Minimum \ac{ul} \ac{se} for different lengths of the square service area $d$; c) CDF of the \ac{ul} per-user \ac{se} in a dense network ($d=200$m, $p_{\max} = 20$ mW).}
\label{fig:results_v1}
\end{figure*}
\begin{figure*}[t!]
\centering
\hfil
\subfloat[]
{\includegraphics[width=0.33\textwidth]{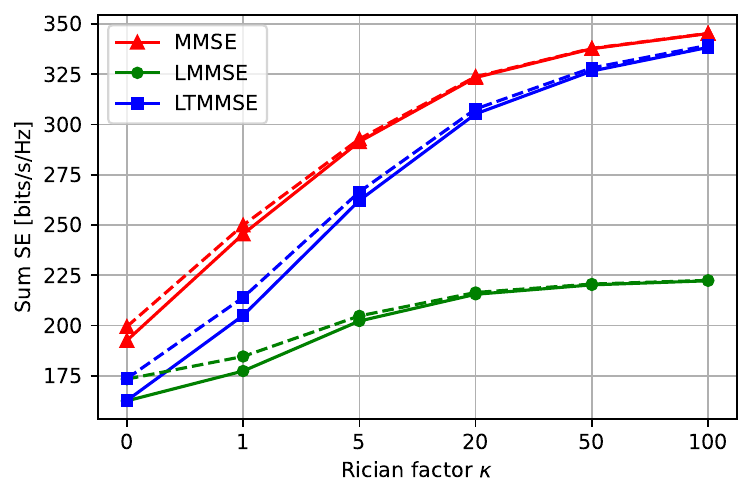}\label{fig:result_k_sumse}}
\hfil
\subfloat[]
{\includegraphics[width=0.33\textwidth]{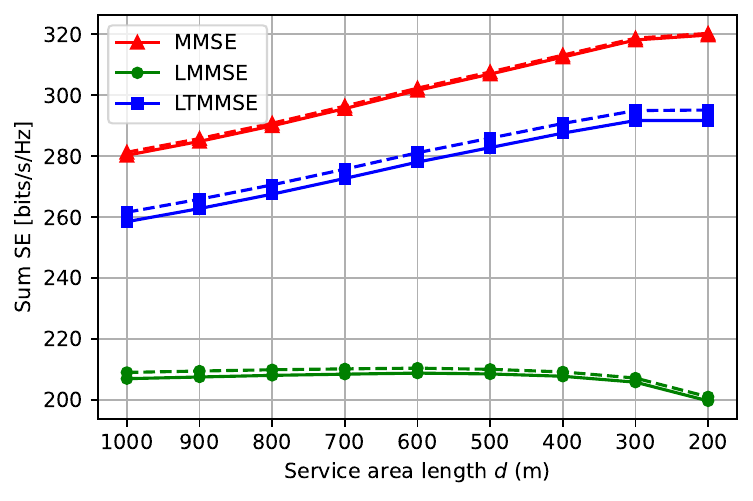}\label{fig:result_d_sumse}}
\hfil
\subfloat[]
{\includegraphics[width=0.33\textwidth]{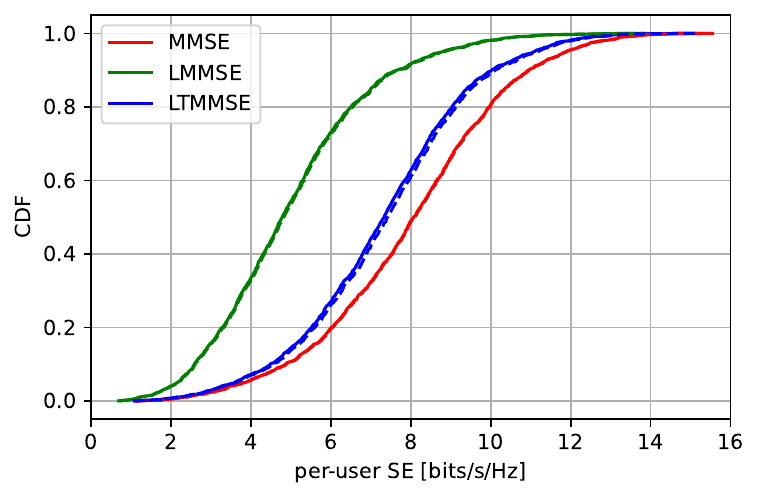}\label{fig:result_cdf_sumse}}
\caption{Comparison of \ac{ul} SEs achieved by different beamforming schemes for the case $v=0$ in \eqref{eq:power_control}. The solid lines refer to the \ac{uatf} bound \eqref{eq:se_uatf}, and the dotted lines refer to the coherent decoding bound \eqref{eq:se_cd}. a) Sum \ac{ul} \ac{se} for different values of $\kappa$ ($d=1$ km, $p_{\max} = 100$ mW); b) Sum \ac{ul} \ac{se} for different lengths of the square service area $d$; c) CDF of the \ac{ul} per-user \ac{se} in a dense network ($d=200$m, $p_{\max} = 20$ mW).}
\label{fig:results_v0}
\end{figure*}

We simulate the performance of the centralized and distributed beamforming schemes covered in Sect. \ref{sec:beamforming} by considering $L=100$ \acp{ap}, each equipped with $N=4$ antennas, and $K=40$ single-antenna \acp{ue}, independently and uniformly distributed over a squared service area of $1000$ m × $1000$ m. A  wrap-around technique is introduced to emulate an infinite service area. All AP-UE pairs have a height difference of $11$ m. The path-loss is computed based on the COST 231 Walfish-Ikegami model for Urban microcell (UMi) scenario \cite[Sect.~5.2]{3gpp}. Specifically, the path-loss $\beta_{k,l}$ between \ac{ap} $l$ and \ac{ue} $k$ is given by
\begin{equation*}
\beta_{k,l}= 35.4 - 20\log_{10}(f_c)-26\log_{10}\left( \frac{d_{k,l}}{1 m}\right)+F_{k,l} \text{ [dB]},
\end{equation*}
where $F_{k,l}\sim\mathcal{N}(0,\sigma_{\text{sf}}^2)$ represents the shadow fading, $f_c$ is the carrier frequency and $d_{k,l}$ is the 3D distance between \ac{ap}~$l$ and \ac{ue}~$k$. By specializing \eqref{eq:rician}, we let $
    \bm{h}_{k,l} =\sqrt{\beta_{k,l}}\left(\sqrt{\frac{\kappa_{k,l}}{\kappa_{k,l}+1}}\bar{\bm{g}}_{k,l}e^{j\theta_{k,l}}+\sqrt{\frac{1}{\kappa_{k,l}+1}}\tilde{\bm{g}}_{k,l}\right)$,
where $\bar{\bm{g}}_{k,l}$ and $\tilde{\bm{g}}_{k,l}\sim\mathcal{N}_{\mathbb{C}}(\bm{0},\bm{R}'_{k,l})$ correspond to \ac{los} and \ac{nlos} components, respectively. The $n$th element of $\bm{\bar{g}}_{k,l}$ is given by $[\bm{\bar{g}}_{k,l}]_n=e^{j2\pi(n-1)\frac{d}{\lambda}\sin(\bar{\phi}_{k,l})\cos(\bar{\varphi}_{k,l})}$, where $\bar{\phi}_{k,l}$ and $\bar{\varphi}_{k,l}$ denote the azimuth angle and the angle of elevation between \ac{ap} $l$ and \ac{ue} $k$, respectively, and $d$ denotes the antenna spacing. We use the Gaussian local scattering model from \cite{9336188} to calculate the spatial correlation matrix $\bm{R}_{k,l}'$. The $(x,y)$th element of $\bm{R}_{k,l}'$ is calculated as
% \begin{equation*}
% \int_{\bar{\varphi}_{k,l}-40^\circ}^{\bar{\theta}_{k,l}+40^\circ}\int_{\bar{\phi}_{k,l}-40^\circ}^{\bar{\phi}_{k,l}+40^\circ} e^{j2\pi\frac{d}{\lambda}(x-y)\sin(\phi)\cos(\theta)}  f_{k,l}(\phi, \theta) \, d\phi \, d\theta,
% \end{equation*}
\begin{equation*}
[\bm{R}_{k,l}']_{x,y} =\int_{-\pi}^\pi \int_0^\pi e^{j2\pi\frac{d}{\lambda}(x-y)\sin(\phi)\cos(\varphi)}  f_{k,l}(\phi, \varphi) \, d\phi \, d\varphi,
\end{equation*}
where $f_{k,l}(\phi, \varphi)$ is the joint probability density function (PDF) of the angles of the multipath components between \ac{ap}~$l$ and \ac{ue}~$k$. As in \cite{9336188}, we assume a jointly Gaussian PDF with mean $(\bar{\phi}_{k,l},\bar{\varphi}_{k,l})$ and diagonal covariance matrix with standard deviations $\sigma_{\phi}=\sigma_{\varphi}=5^{\circ}$, truncated to $8$ standard deviations, wrapped around the angular support, and renormalized. We refer to \cite[Sect.~2.5.3]{9336188} for additional details. The Rician factor $\kappa_{k,l}$, which models the relative strength of the \ac{los} component, is calculated according to 3GPP specifications \cite{3gpp}
\begin{equation}\label{eq:kappa}
\kappa_{k,l}=10^{1.3-0.003d_{k,l}}.
\end{equation}
For the pilot assignment and formation of user-centric cooperation clusters in the network, we use the sequential \ac{dcc} algorithm proposed in \cite[Algorithm~4.1]{9336188}. The \ac{ul} transmit power for each \ac{ue} $k\in\mathcal{K}$ is determined using the fractional power control formula \cite[Eq.~(7.34)]{9336188}
\begin{equation}\label{eq:power_control}
p_k = p_{\text{max}} \frac{\left(\sum_{l\in\mathcal{L}_k}{\beta_{k,l}}\right)^{v}}{\max_{i\in\{1,\dots,K\}}\left(\sum_{l\in\mathcal{L}_i}{\beta_{i,l}}\right)^{v}},
\end{equation}
where we consider the two cases $v = -1$ and $v = 0$. The case $v = -1$ let $p_k\sum_{l\in\mathcal{L}_k}{\beta_{k,l}}$ be identical for all \acp{ue}, and it approximates a max-min fair power control policy \cite{9336188}. The case $v = 0$ let all \acp{ue} transmit with the same power $p_{\text{max}}$, and it approximates a sum-SE optimal power control policy \cite{9336188}. Based on the network requirements, these policies can be used to optimize either the total system performance or the individual user experience.

\subsection{Results and Conclusions}

In Fig. \ref{fig:results_v1} we first focus on the (approximate) max-min fair power control policy. Fig. \ref{fig:result_k_minse} illustrates the impact of \ac{los} on the relative performance of different beamforming schemes covered in Sect. \ref{sec:beamforming}. It plots the minimum \ac{ul} \ac{se} for different values of a common Rician factor $(\forall k \in \mathcal{K})(\forall l \in \mathcal{L})$ $\kappa_{k,l}= \kappa \in[0,100]$. In agreement with Proposition~\ref{prop:equivalence}, for $\kappa=0$, representing a \ac{nlos} channel, the minimum \ac{ul} \ac{se} is the same for the \ac{ltmmse} and \ac{lmmse} schemes, with the \ac{mmse} scheme significantly outperforming both distributed schemes. As the value of $\kappa$ increases, indicating a transition towards stronger \ac{los} conditions,  the \ac{ltmmse} scheme begins to outperform the \ac{lmmse} scheme. This is particularly noticeable for $\kappa=5$ or higher. Eventually, the performance of the \ac{ltmmse} scheme converges to that of the \ac{mmse} scheme, in agreement with Proposition~\ref{prop:equivalence2}.  
 
We then assess the performance for different network densities, by varying the length $d$ of the square service area while maintaining a constant number of \acp{ap} and \acp{ue} as detailed in Table~\ref{tab:parameters}. We decrease the maximum \ac{ul} transmit power $p_{\text{max}}$ from $100$~mW to $20$~mW proportionally to $d$, to ensure a fair comparison in terms of \ac{snr} across various network densities. Fig. \ref{fig:result_d_minse} shows the minimum \ac{ul} \ac{se} for different network densities. As the service area shrinks, $\kappa_{k,l}$ in  \eqref{eq:kappa} increases, resulting in stronger \ac{los} components, i.e., in a stronger channel mean. As expected, the performance of the \ac{ltmmse} scheme %beamforming optimally exploits the statistical properties of the channel, and its achieved minimum \ac{ul} \ac{se} 
increases significantly along with the performance of the \ac{mmse} scheme for denser networks. In contrast, \ac{lmmse} doesn't exhibit a similar improvement in performance, which is possibly because of its inability to handle well the interference originating from the strong \ac{los} components. In Fig. \ref{fig:result_cdf_minse}, we plot the CDF of the SEs achieved by different beamforming schemes for the densest network from our simulations. It can be noticed that the \ac{ul} \ac{se} achieved by \ac{ltmmse} beamforming approaches the UL SE achieved by \ac{mmse} beamforming.

In Fig.~\ref{fig:results_v0}, we repeat the above analysis by investigating the (approximate) sum-SE optimal power control policy. The results indicate similar trends as for the minimum rate case, except for a somewhat counter-intuitive decrease in the sum-SE as the network density exceeds a certain threshold in Fig. \ref{fig:result_d_sumse}. This decrease is likely due to the suboptimal power control policy, which leads to excessive interference dense setups. We predict that a sum-SE optimal power control policy would remove some users from service and ensure a consistent sum-SE growth as the network density increases. However, this is challenging to verify experimentally, since we recall that sum-SE optimal power control is known to be NP-hard.

Finally, we study the impact of using the different \ac{se} lower bounds in Sect. \ref{sec:data_transmission}. It can be seen from both Fig.~\ref{fig:results_v1} and Fig.~\ref{fig:results_v0}  that, for all experiments, the \ac{uatf} bound \eqref{eq:se_uatf} is an excellent proxy for optimizing the more accurate yet intractable coherent decoding bound \eqref{eq:se_cd}. This is particularly evident for large $\kappa$  values in Fig. \ref{fig:result_k_minse}. This can be explained by the fact that for $\kappa\to\infty$, all channels are deterministic and hence the expectations in \ac{uatf} and coherent decoding bound can be removed, and the two bounds coincide.

% if have a single appendix:
% \appendix[Proof of the Zonklar Equations]
% or
% \appendix  % for no appendix heading
% do not use \section anymore after \appendix, only \section*
% is possibly needed

% use appendices with more than one appendix
% then use \section to start each appendix
% you must declare a \section before using any
% \subsection or using \label (\appendices by itself
% starts a section numbered zero.)

% \appendices
% \section{Proof of the First Zonklar Equation}
% Appendix one text goes here.

% % you can choose not to have a title for an appendix
% % if you want by leaving the argument blank
% \section{}
% Appendix two text goes here.

% % use section* for acknowledgment
% \section*{Acknowledgment}

% The authors would like to thank...

% Can use something like this to put references on a page
% by themselves when using endfloat and the captionsoff option.
\ifCLASSOPTIONcaptionsoff
  \newpage
\fi

% trigger a \newpage just before the given reference
% number - used to balance the columns on the last page
% adjust value as needed - may need to be readjusted if
% the document is modified later
%\IEEEtriggeratref{8}
% The "triggered" command can be changed if desired:
%\IEEEtriggercmd{\enlargethispage{-5in}}

% references section

% can use a bibliography generated by BibTeX as a .bbl file
% BibTeX documentation can be easily obtained at:
% http://mirror.ctan.org/biblio/bibtex/contrib/doc/
% The IEEEtran BibTeX style support page is at:
% http://www.michaelshell.org/tex/ieeetran/bibtex/
\bibliographystyle{IEEEbib}
\bibliography{bibtex/bib/references}

\end{document}